\documentclass
[12pt]
{article}
\usepackage{amsmath}
\usepackage{amsfonts}
\usepackage{graphicx}
\newcommand{\bean}{\begin{eqnarray*}}
\newcommand{\eean}{\end{eqnarray*}}
\newcommand{\ed}{\end{document}}
\newcommand{\bea}{\begin{eqnarray}}
\newcommand{\eea}{\end{eqnarray}}

\begin{document}
\title{A new approach to
the Baker-Campbell-Hausdorff expansion
}
\author{A.V.Bratchikov
\\ Kuban
State Technological University,\\ 2 Moskovskaya Street, Krasnodar,
  350072, Russia
\\
}
 \date{} \maketitle
\begin{abstract} 
For noncommutative variables $x,y$ an expansion of  $\log(e^{x}e^{y})$ in powers of $x+y$ is obtained.
Each term of the series is given by an infinite sum in powers of $x-y.$
The series is represented by diagrams. 
\end{abstract}


\bigskip

Let $L$ be a Lie algebra and let $f(x,y)$ be defined by 
$$f(x,y)=\log(e^{x}e^{y}).$$
The expansion of $f(x,y)$ in  powers of $x,y$ 
can be represented in the form
\bea \label{1}f(x,y)=\sum_{m=1}^\infty p_m(x,y),
\eea
where $p_m(x,y)$ is a homogeneous polynomial of degree $m$ in $x$ and $y.$
Every polynomial $p_m(x,y)$ can be expressed as a sum of $x,y$ and nested commutators of $x$ and $y.$ 
Equation (\ref{1}) is known as the Baker-Campbell-Hausdorff (BCH) series. 
An explicit formula for $p_m(x,y)$ was derived by Dynkin \cite{D}.

The BCH series plays an important role in theoretical physics (see e.g. 
\cite {BL},
\cite {Z}). Recent applications are connected with deformation quantization of Poisson manifolds \cite {K}, \cite{Br}.

For $x+y=0$ we get $f(x,y)=0,$ and therefore $f(x,y)$ can be represented in the form 
\bea \label {blya}
f(x,y)=\sum_{m=1}^\infty q_m(x-y,x+y),
\eea
where $q_m(x-y,x+y)$ is a homogeneous polynomial of degree $m$ in $x+y.$
This equation 
can be also written  as 
\bean 
f((x+y)/ 2,(x-y)/ 2)=\sum_{m=1}^\infty q_m(y,x).
\eean
The aim of this paper is to find an expression for $q_m$ and construct 
a graphical representation of the series in terms of diagrams.

For $x,y\in L$ put 
$$\psi(t,x,y)= \log(e^{tx}e^{ty}),\qquad t\in \mathbb R.$$ Let $k(z)$ be the function on $\mathbb C$ given by  $k(z)= z/(1-e^{-z}) - z/2.$ One can show \cite{M} that 
\begin{eqnarray} \label {qu}
k(z)=1+\sum_{p=1}^\infty k_{2p}z^{2p},
\end{eqnarray}
where $k_{2p}=B_{2p}/(2p)!,$ and $B_{2p}$ are the Bernoulli numbers
$$B_2=\frac 1 6,\quad B_4=- \frac 1 {30},\quad B_6= \frac 1 {42},\ldots.$$
Then $\psi(t,x,y)$   
is a solution of the equation \cite{V} 
\begin{eqnarray}  \label{As}
\frac {d\psi}{dt}=k(ad \psi)(x+y)+ \frac 1 2 [x-y,\psi]
\end{eqnarray}
with the initial condition 
$$\psi (0,x,y)=0.$$

Equation (\ref{As}) is easily solvable with the solution given implicitly by
\bea  \label{A}
\psi(t)=e^{\frac t 2 ad(x-y)} 
\int_{0}^t e^{-\frac \tau 2 ad(x-y)}k(ad\psi(\tau))(x+y)d\tau
.
\eea
This can be written in the form 
\bea \label{bas}
\psi =\psi_0+\sum_{n=1}^\infty g_{2n}(\psi),
\eea
where
\bean  \label{b}\psi_0=e^{\frac t 2 ad(x-y)} 
\int_{0}^t e^{-\frac \tau 2 ad(x-y)}(x+y)d\tau=
\sum_{n=0}^\infty \frac {t^{n+1}} {(n+1)!2^n} \left(ad(x-y)\right
)^n(x+y)
,
\eean
$$g_{2n}(\psi)=k_{2n}e^{\frac t 2 ad(x-y)} 
\int_{0}^t e^{-\frac \tau 2 ad(x-y)}\left(ad\psi(\tau)\right)^{2n}(x+y)d\tau
.$$

Let us introduce the 
functions 
\begin{eqnarray*} \langle \ldots.  \rangle_{2n}:
{L }^{2n} 
 \to L,\quad n=1,2,\ldots,
 \end{eqnarray*} defined for $v_1,\ldots,v_{2n}\in L$ by 
\begin{eqnarray} \label {osk}
\langle v_1,\ldots,v_{2n} \rangle_{2n} = 
\frac \partial {\partial\alpha_1}\ldots \frac \partial {\partial\alpha_{2n}}g_{2n}( \alpha_1v_{1}+\ldots+ \alpha_{2n}v_{2n}).
\end{eqnarray} 
The function  $\langle v_1,\ldots,v_{2n} \rangle_{2n}$ is linear separately in each $v_i,$ and symmetric.
From (\ref{osk}) it follows
\begin{eqnarray*} 
\langle v,\ldots,v \rangle_{2n}  =  {(2n)!} g_{2n}(v). 
\end{eqnarray*} 
Then equation (\ref{bas}) takes the form 
\begin{eqnarray} \label {equv}
\psi= k \psi_0+k\sum_{n=1}^\infty \frac 1 {(2n)!} \langle \psi,\ldots,\psi \rangle_{2n}.
\end{eqnarray}
Here we introduced an auxiliary parameter $k$ for counting powers of $x+y.$
Eventually it is set to be $1.$

To describe a solution of  equation (\ref {equv}) we introduce a family of functions 
$$P^m_{i_1\ldots i_{2n}}: L^m\to L^{m-2n+1},$$
$m\geq 2,1\leq i_1< \ldots < i_{2n}\leq m,$
defined by
\begin{eqnarray*} \label {or}
P^m_{i_1\ldots i_{2n}}( v_1,\ldots ,v_m ) = 
(h \langle v_{i_1},\ldots,{v}_{i_{2n}}\rangle_{2n} ,v_1,\ldots,\widehat{v}_{i_1},\ldots,\widehat{v}_{i_{2n}},\ldots,v_{m} ),
\end{eqnarray*} 
where $\widehat{v}$ means that ${v}$ is omitted.
If $v\in L$ is given by  
\begin{eqnarray} \label {v} v = P^{n_s}_{I_s}\ldots P^{m-n_1+1}_
{I_2}
P^{m}_{I_1}
(v_1,\ldots ,v_m )
\end{eqnarray}
for some $I_1= (i_1^1,\ldots,  i^1_{n_1}),
I_2=(i^2_1,\ldots, i^2_{n_2}),
\ldots,I_s=(i^s_1,\ldots, i^s_{n_s}), n_1+\ldots+n_s-s+1=m,
$ we say that  $v$ is a descendant of $(v_1,\ldots ,v_m ).$
A descendant of $v$ is defined as $v.$

Each  descendant can be represented by a diagram. 
In this diagram an element of  $L$ is represented by the line segment \rule[3pt]{20pt}{0.5pt}\,\,. A product $$(v_1,\ldots,v_{2n})\to k\langle v_1,\ldots,v_{2n}\rangle_{2n}$$ is represented by the vertex joining the line segments for  $ v_1,\ldots,v_{2n},$ $ k\langle v_1,\ldots,v_{2n}\rangle_{2n}.$ 
The vectors  $v_1,\ldots ,v_m$ are called incoming.
The function $ k\langle v_1,\ldots,v_{2n}\rangle_{2n}$ is called the vertex function.
The general rule for graphical representation of $P^{m}_{I}
(v_1,\ldots ,v_m )$ should be clear
from Figure~1.Here we show the diagram for $$P^m_{ij}(v_1,\ldots,v_m)=(k \langle v_{i},v_j\rangle_2 ,v_1,\ldots,\widehat{v}_{i},\ldots,\widehat{v}_{j},\ldots,v_{m} ).$$
The points labeled  by $1,\ldots, m$ represent the ends of the lines for $v_1,\ldots ,v_m.$ Using the representation for $P^m_{i_1\ldots i_{2n}}( v_1,\ldots ,v_m )$ one can consecutively draw the diagrams for $P^{n_1}_{I_1}(v_1,\ldots ,v_m ),P^{n_2}_{I_2}
P^{n_1}_{I_1}(v_1,\ldots ,v_m ),\ldots , v$ (\ref {v}). The vector $v$ is called outgoing.

\begin{figure}
\centering
\resizebox{0.35\textwidth}{!}{%
  \includegraphics{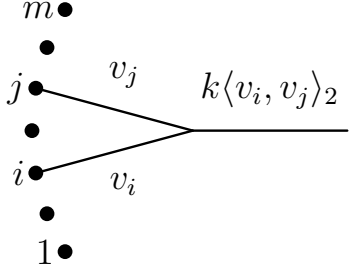}
}
\caption{ Diagram for $P^m_{ij}(v_1,\ldots,v_m).$ }
\end{figure}


Let us introduce a family of functions $$\langle \ldots  \rangle:
{L }^m 
 \to L,\qquad m=1,2,\ldots,$$
such that for $v_1,\ldots,v_m \in V$ $\langle v_1,\ldots,v_m \rangle$ is 
defined as the sum of all the descendants of its arguments.
For example, 
\begin{eqnarray*} \label {or}
\langle v_1,v_2 \rangle = k\langle v_{1},v_{2}\rangle_2,
\end{eqnarray*} 
\begin{eqnarray*} \label {or}
\langle v_1,v_2,v_{3}\rangle=k^2 \left(
\langle \langle v_{1},v_{2}\rangle_2, v_3 \rangle_{2}+\langle \langle v_{1},v_{3}\rangle_2, v_2 \rangle_{2}+\langle \langle v_2, v_3 \rangle_2, v_{1}\rangle_2\right). 
\end{eqnarray*}

A solution of equation (\ref {equv}) is given by \cite{B1}
\bea \label {oo}
\psi= \langle e^{k \psi_0} \rangle, 
\eea
where
\bean \label {}
\langle e^{k \psi_0} \rangle = \sum_{n=0}^\infty \frac {k^n} {n!}\langle 
\psi_0^n \rangle
,
\eean
$\langle 
\psi_0^n\rangle$ is defined as $0$ if $n=0,$ and otherwise $$\langle 
\psi_0^n\rangle=\langle 
\underbrace{\psi_0,\ldots ,\psi_0}_{\text {n times}}\rangle
.$$

We write 
\bea \label {oav}
\psi= \sum_{n=1}^\infty 
\pi_n(t,x-y,x+y). 
\eea
Here $\pi_n
$ is a homogeneous polynomial of degree $n$ in $x+y.$
The functions $q_m$ (\ref {blya}) are defined by $$q_m(x-y,x+y)=\pi_m(1,x-y,x+y).$$
Using (\ref{oo}), we get
\bean 
\pi_1= k\psi_0,\quad \pi_3=\frac {k^3}{24}  e^{\frac t 2 ad(x-y)} 
\int_{0}^t e^{-\frac \tau 2 ad(x-y)}\left(ad\psi_0(\tau)\right)^{2}(x+y)d\tau,\quad \pi_2=\pi_4=0.
\eean
The function $\pi_n(x-y,x+y)$ is represented by a sum of diagrams.It follows from (\ref{oo}) that each incoming line contributes a factor of $k$ to a such diagram,
and therefore 
$\pi_n$ 
is given by a sum of the diagrams with $r$ incoming lines and $s$ vertices such that $r+s=n.$ 

For example, the $O(k^5)$ contribution in $\psi$ is given by  
\begin{eqnarray*}
\pi_5=\frac {k^5} 2
\langle\langle \psi_0, \psi_0\rangle_2, \psi_0\rangle_2+\frac {k^5} {24}\langle 
\psi_0,\psi_0
, \psi_0,
\psi_0\rangle_4.
\end{eqnarray*}
The diagram for $k^2\langle\langle \psi_0, \psi_0\rangle_2, \psi_0\rangle_2 $ is depicted in 
Figure 2. 

\begin{figure}
\centering
\resizebox{0.60\textwidth}{!}{%
  \includegraphics
  {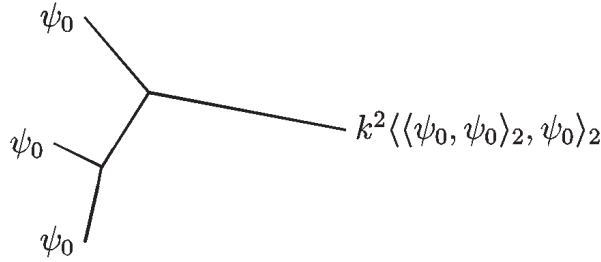}
}
\caption{ Diagram for $k^2\langle\langle \psi_0, \psi_0\rangle_2, \psi_0\rangle_2.$}
\end{figure}



\bigskip

\end{document}